\pdfoutput=1
\documentclass{article}

\usepackage{icrctc07}

\newlength{\figurewidth}
\title{A radio air shower surface detector as an extension for IceCube and
  IceTop}

\shorttitle{A radio air shower surface detector as an extension for
  IceCube and IceTop}

\authors{J. Auffenberg$^1$, T. Gaisser$^2$, K. Helbing$^1$,
  T. Huege$^3$, T. Karg$^1$, A. Karle$^4$}

\shortauthors{J. Auffenberg et al.} 

\afiliations{$^1$Bergische Universit\"at Wuppertal, Fachbereich C --
  Astroteilchenphysik, 42097 Wuppertal, Germany \\
  $^2$Bartol Research Institute, University of Delaware, Newark, DE
  19716, USA \\
  $^3$Institut f\"ur Kernphysik, Forschungszentrum Karlsruhe, Postfach
  3640, 76021 Karlsruhe, Germany \\
  $^4$Department of Physics, University of Wisconsin, Madison, WI
  53706, USA}

\email{jauffenb@uni-wuppertal.de}

\abstract {The IceCube neutrino detector is built into the Antarctic
  ice sheet at the South Pole to measure high energy neutrinos. For
  this, 4800 photomultiplier tubes (PMTs) are being deployed at depths
  between 1450 and 2450 meters into the ice to measure neutrino
  induced charged particles like muons. IceTop is a surface air shower
  detector consisting of 160 Cherenkov ice tanks located on top of
  IceCube. To extend IceTop, a radio air shower detector could be
  built to significantly increase the sensitivity at higher shower
  energies and for inclined showers.  As air showers induced by cosmic
  rays are a major part of the muonic background in IceCube, IceTop is
  not only an air shower detector, but also a veto to reduce the
  background in IceCube. Air showers are detectable by radio signals
  with a radio surface detector. The major emission process is the
  coherent synchrotron radiation emitted by $e^+$ $e^-$ shower
  particles in the Earths magnetic field (geosynchrotron effect).
  Simulations of the expected radio signals of air showers are shown.
  The sensitivity and the energy threshold of different antenna field
  configurations are estimated.}
  
\begin{document}
\maketitle

\section{Introduction}

\begin{figure}[ht]
  \centering
  \includegraphics[width=\figurewidth]{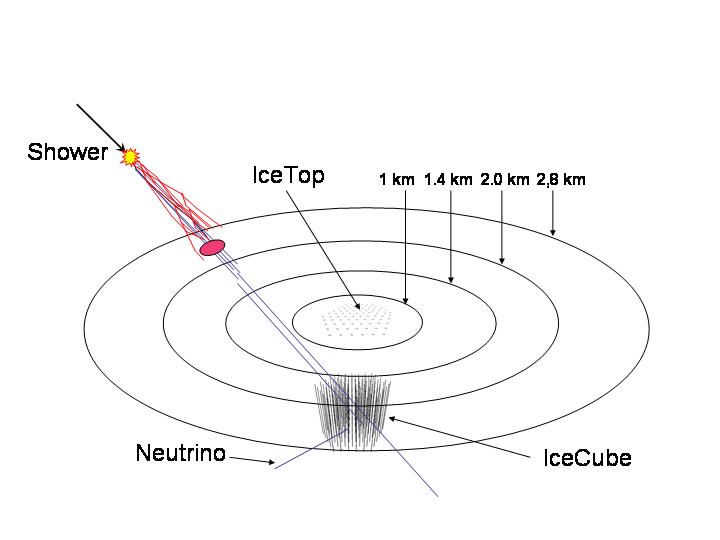}
  \caption{Schematic view of an antenna array together with IceTop
    (hexagon) and IceCube. The surface antennas would be installed on
    the rings. At higher distances only more inclined showers can hit
    IceCube for which simulations predict stronger radio signals.
    This allows larger antenna spacing for the outer rings.}
  \label{fig01}
\end{figure}

When the IceCube\footnote{http://icecube.wisc.edu/} neutrino telescope
will be finished in 2011 it will consist of up to 80 strings deployed
in the Antarctic ice, each one containing 60 PMTs (Fig.~\ref{fig01}).
One of the main goals of IceCube is to measure high energy neutrinos
from cosmic sources.  It is designed to measure up-going neutrino
induced muons, since only neutrinos are not absorbed in the Earth.

IceTop is built on the surface above IceCube (Fig.~\ref{fig01}) to
detect air showers in the energy range from 10$^{15}$\,eV to
10$^{18}$\,eV and to study the composition of cosmic rays.

The emission of coherent synchrotron radiation from $e^+$ $e^-$ shower
particles in the Earth magnetic field can be measured \cite{CGR} and
could be the basis for an extension of the IceTop and IceCube
detection systems.

\section{Possible Radio Air Shower Arrays}

The LOPES collaboration has shown air shower detection by its radio
emission and demonstrated the possibility to reconstruct the
inclination angle and the shower core from the radio signal
\cite{LOPES}. There are mainly two ways that a radio air shower surface
detector for IceCube could improve IceTop and IceCube.

\begin{figure}[ht]
  \centering
  \includegraphics[width=\figurewidth]{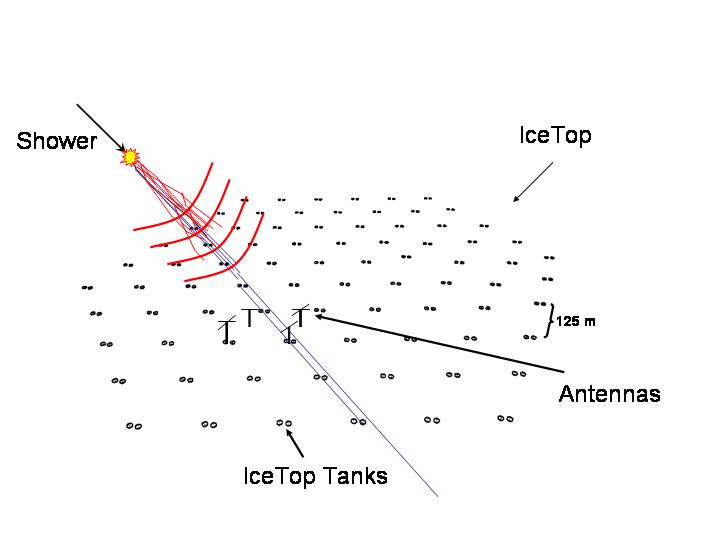}
  \caption{Possible antenna field configuration that could serve for
    first test measurements.  An infill array would look similar. The
    black dots are the Ice Top tank positions. The antenna positions
    of an infill array are in-between the tanks. A test array should
    be on top and in-between the Ice Top tanks to determine the RFI
    from Ice Top.}
  \label{fig02}
\end{figure}

\subsection {Expansion of the Surface Array}

An expansion of the area of IceTop with surface radio antennas would
be an enhancement of the air shower detector to higher primary
energies and to more inclined showers. The idea is to build an antenna
array in rings of increasing radius around the IceTop array. Muon
bundles from air showers can be mis-reconstructed as an up-going
signal in the IceCube detector. With an expansion in area of the
IceTop detector it would be possible to detect and veto signals from
very inclined air showers (cf.~Fig.~\ref{fig01}).

\subsection {Infill Surface Array}

An infill surface radio detector could be built on the same area as
IceTop with similar distances, but shifted positions with respect to
the Cherenkov tank array. This would provide an additional powerful
observation technique for cosmic ray research of the same showers. In
addition the radio detector field could be an upgrade to higher
inclination angles. The infill array would allow one to detect high
energy air showers with three independent detector systems: IceTop,
IceCube, and the radio surface array (Fig.~\ref{fig02}).

\section{First Background Studies}

\begin{figure}[ht]
  \centering
  \includegraphics[width=\figurewidth]{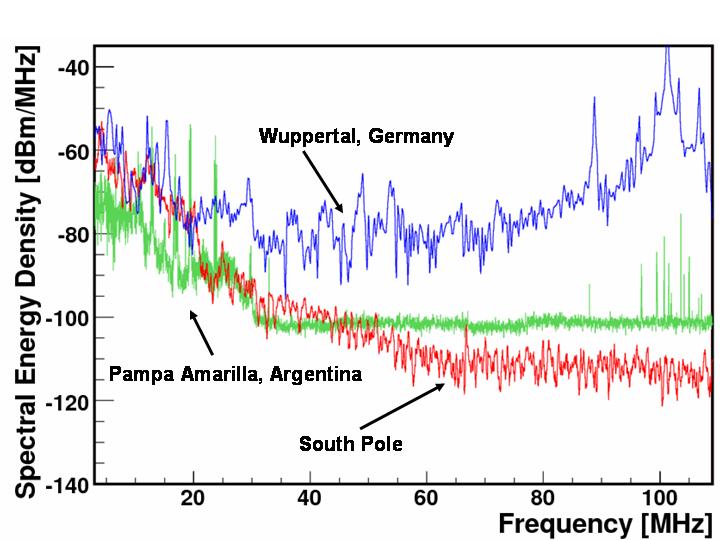}
  \caption{Background studies with the 3\,m monopole antenna in
    Wuppertal (Germany), in the Pampa Amarilla (Argentina), and at the
    South Pole.  The Wuppertal and South Pole spectra are Fourier
    transforms of time sweeps recorded with a digital oscilloscope.
    The Argentina spectrum is measured with a spectrum analyzer.}
  \label{fig03}
\end{figure}

First studies of the radio background at the South Pole have been
carried out in November 2006 with a 3\,m monopole antenna located
close to the AMANDA counting house (MAPO). The signal was amplified
using a 39\,dB commercial pre-amplifier
(MITEQ\footnote{http://www.miteq.com/} AU-1464), transmitted over
100\,m RG-58 cable, and recorded with a digital oscilloscope stationed
in the MAPO building. The Discrete Fourier Transform of the time
sweeps gives the spectral energy density of the radio background. It
is corrected for the frequency independent gain of the pre-amplifier
and for a mean antenna gain based on calculations with the antenna
simulation program EZNEC\footnote{http://eznec.com/}. The antenna
amplification was averaged over the half solid angle, and the
polarization of the background signal was assumed to be isotropic.
Figure \ref {fig03} shows the radio background for different regions
measured with the same antenna-amplifier set-up. Figures \ref{fig04}
to \ref{fig06} show the measured background together with simulated
air shower signals. The high background below 20\,MHz has to be
investigated in more detail in larger distances to the South Pole
Station. At higher frequencies the background drops to
$-$110\,dBm/MHz. The spectrum measured in Argentina seems to be lower
than the South Pole spectrum at frequencies below 20\,MHz but higher
at larger frequencies. The reason might be RFI form the AMANDA
counting house at the South Pole. Above 80\,MHz the Argentina spectrum
shows sharp monofrequent radio transmitters. The spectrum at Wuppertal
is higher at all frequencies, as Wuppertal is a region with intense
noise from civilization.  There are only few mono-frequent
transmitters in the region from 5\,MHz to 120\,MHz at the South Pole.
So the background at the South Pole will allow for broad-band
measurements of radio air shower signals.

\section{Radio Air Shower Simulation}

The REAS2 Code \cite{reas2}, based on air showers simulated with
CORSIKA \cite{CORSIKA}, is used to evaluate the electric field
vector of the geosynchrotron emission of air showers at the South
Pole. We use proton induced air showers with primary energies ranging
from 10$^{16}$\,eV to 10$^{18}$\,eV and South Pole atmosphere and
magnetic field.

For the simulation of the radio emission, the local magnetic field
$\vec{B}$ and the observation height (2800\,m) are important. At the
South Pole the field strength is $\vert \vec{B} \vert =$ 55.3\,$\mu$T
and the field inclination angle is $\Phi = -$72.6$^\circ$
\cite{Bfield}.  Since the power density is given by the norm of the
Pointing vector $\vert \vec{S} \vert = \frac{1}{Z_{0}} \, \vec{E}^{2}$
with $Z_{0} = \sqrt{\mu_0/\epsilon_0}$ the spectral electric field
density calculated in the simulation can be easily converted to
spectral power density.

\begin{figure}[ht]
  \centering
  \includegraphics[width=\figurewidth]{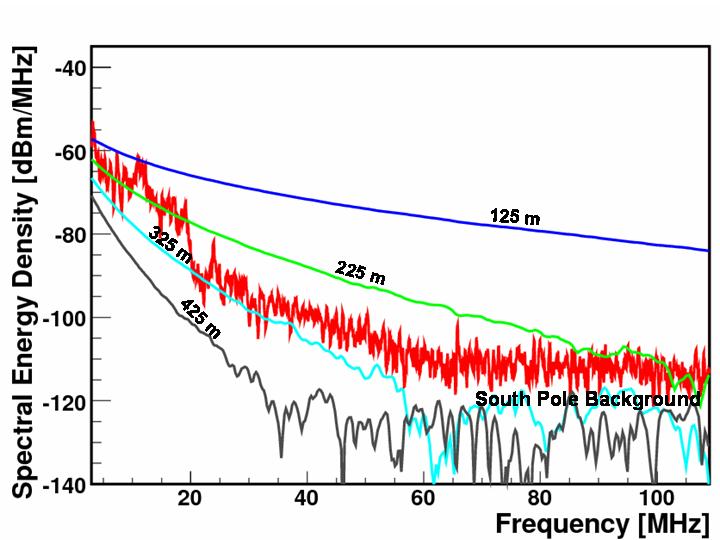}
  \caption{Radio emission from a proton induced 10$^{17}$\,eV air
    shower with inclination angle $\theta =$ 20$^\circ$ for antennas
    at different distances to the shower core position on the ground.
    For comparison radio background measurements from the South Pole
    are shown (red curve).}
  \label{fig04}
\end{figure}

\begin{figure}[ht]
  \centering
  \includegraphics[width=\figurewidth]{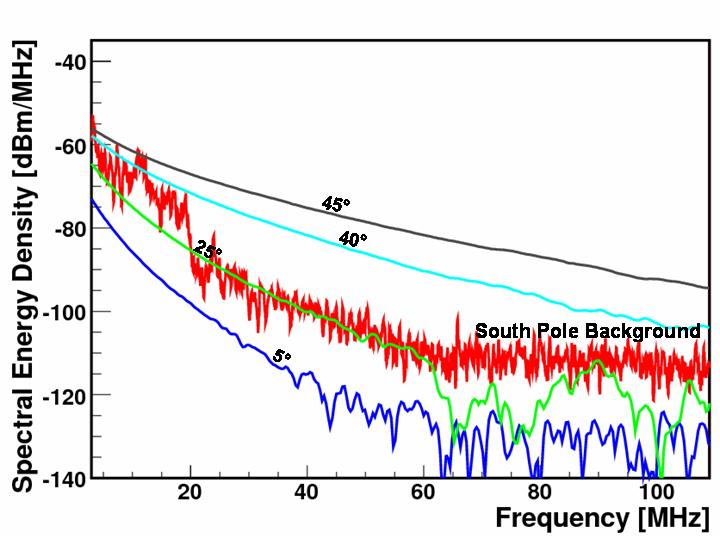}	
  \caption{Radio emission from a proton induced 10$^{17}$\,eV air
    shower with 325\,m distance to the shower core position on the
    ground for different inclination angles from $\theta =$ 5$^\circ$
    to 45$^\circ$. For comparison radio background measurements from
    the South Pole are shown (red curve). The highest line is
    the same as in Figure \ref{fig06}.}
  \label{fig05}
\end{figure}

\begin{figure}[ht]
  \centering
  \includegraphics[width=\figurewidth]{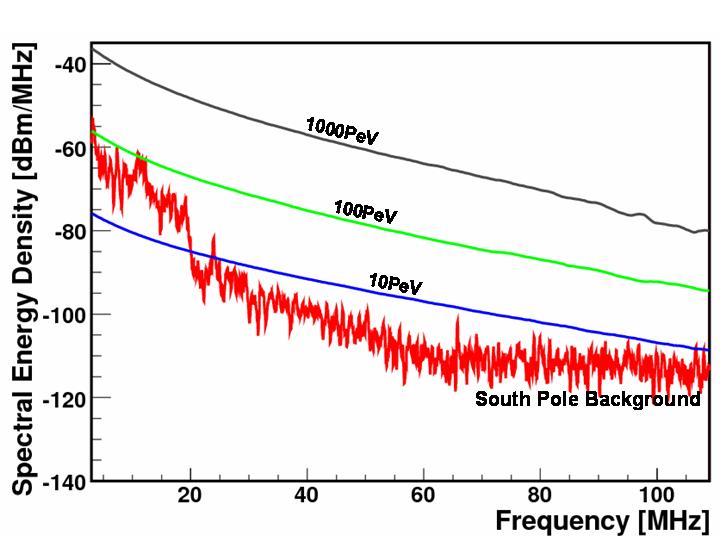}	
  \caption{Radio emission from a proton induced shower with an
    inclination angle $\theta =$ 45$^\circ$ at 325\,m distance to the
    shower core position on the ground for different shower energies
    from 10$^{16}$\,eV to 10$^{18}$\,eV shown together with background
    measurements from the South Pole. Even lower energetic showers
    will be detectable at higher inclination angles.}
  \label{fig06}
\end{figure}

Results of the REAS2 simulation are shown in Figures \ref{fig04} to
\ref{fig06}. Figure \ref{fig04} shows, that a 10$^{17}$\,eV proton
induced air shower with an inclination angle $\theta =$ 20$^\circ$
from 290$^\circ$W should be detectable by antennas with distances up
to about 300\,m from the shower core position on the ground if we
require a signal to background ratio of unity. Figure \ref{fig05}
shows that for more inclined showers the detection range increases.
The radio signal is nearly independent of the azimuthal direction of
the air shower. For higher shower energies the detection sensitivity
increases strongly (Fig.~\ref{fig06}).

One result of comparisons of air shower simulation and background
measurements is, that the distances between the antennas can be larger
for more inclined showers. Thus in the option of expanding IceTop, the
distance between antennas can increase with distance from IceCube. The
range increases with the inclination angle $\theta = \arctan\,(r/d)$
of the showers which could reach the edge of IceCube ($d$ is the depth
of IceCube (2500\,m), and $r$ the distance to the edge of IceCube
projected on the South Pole surface). If we require again a signal to
background ratio of unity for air shower detection one can make the
following predictions.  In 1000\,m distance from the edge of IceCube
the inclination angle of a shower which could reach IceCube is
$\arctan\,($1$/$2.5$) =$ 21.8$^\circ$. To detect the radio signal of a
10$^{17}$\,eV shower with this inclination angle, the antennas have to
be about 300\,m distant.  Right above the core of IceCube, the
distance of the antennas has to be 200\,m to detect vertical showers.

\section{Conclusions and Outlook}

A comparison of background measurements and simulation results shows,
that the lower energy threshold of a radio air shower surface detector
at the South Pole will be around 10$^{17}$\,eV dependent on the
antenna spacing. An infill detector of IceTop with radio antennas
would allow one to cross-calibrate both detection systems. The
expansion in area is a promising possibility for vetoing muon bundles
from inclined air showers in IceCube.

To improve predictions concerning the radio signal emitted by air
showers, more detailed simulations will be made up to higher primary energies
and inclination angles. Further, the polarization of the radio
emission will be studied. These results have to be confirmed by events
measured with a test antenna array at the South Pole.

To investigate the properties of the background in more detail a multi
antenna array is planned. Figure \ref{fig02} shows a possible array
configuration to measure at four positions with all field orientations.

The simulation underlines the effectivity of an antenna field for
increasing inclination angles (Fig.~\ref{fig05}). For higher energies
the shower will be detectable at larger distances to the antennas
(Fig.~\ref{fig06}).

The results of radio air shower simulations and background
measurements at the South Pole show that both, the expansion in area
and the radio infill array are excellent detection systems to improve
the existing detection systems IceCube and IceTop.

\paragraph{Acknowledgements} This work is supported by the German
Ministry for Education and Research and the U.S.~National
Science Foundation.


\begin{thebibliography}{99}

\bibitem{CGR}
  T.~Huege, H.~Falcke, Astron. Astrophys. {\bf 412} (2003) 19.

\bibitem{LOPES}
  H.~Falcke et al., Nature {\bf 435} (2005) 313.

\bibitem{reas2} T.~Huege, R.~Ulrich, R.~Engel, Astropart. Phys. {\bf
    27} (2007) 392.

\bibitem{CORSIKA}
  D.~Heck et al., Report {\bf FZKA 6019} (1998), \\
  http://www-ik.fzk.de/corsika/
  physics\_description/corsika\_phys.html.

\bibitem{Bfield}
  National Geophysical Data Center (NGDC),
  http://www.ngdc.noaa.gov/seg/geomag/jsp/ IGRF.jsp.

\end{thebibliography}
\end{document}